\documentclass[aps,pre,twocolumn,groupedaddress,showpacs,amsfonts,amssymb,amsmath,floatfix]{revtex4}
\usepackage{hyperref}
\usepackage{dcolumn}
\usepackage{graphicx}
\usepackage{bm}
\usepackage{natbib}
\usepackage{url}
\usepackage{cancel}

\newcommand{\units}[2]{#1\nobreak\mbox{$\:\mathrm{#2}$}}
\newcommand{\ie}{\textit{i.e.}\ }

\newcommand{\eg}{\textit{e.g.}\ }
\newcommand{\etal}{\mbox{\textit{et al.}}}
\newcommand{\via}{\textit{via}\ }

\newcommand{\ed}{\varepsilon_\text{d}}
\newcommand{\ld}{\kappa^{-1}}
\newcommand{\lb}{\ell_\text{B}}
\newcommand{\lgc}{\ell_\text{GC}}
\newcommand{\rhoe}{\rho_\text{e}}
\newcommand{\rhos}{\rho_\text{s}}
\newcommand{\kt}{k_\text{B}T}

\newcommand{\br}{{\bm r}}

\newcommand{\rhof}{\rho_\text{f}}
\newcommand{\rhoi}{\rho_\text{i}}

\newcommand{\dep}[2]{\frac{\partial #1}{\partial #2}}
\newcommand{\ddep}[2]{\frac{\partial^2 #1}{\partial #2 ^2}}

\begin{document}

\title{Liquid friction on charged surfaces: from hydrodynamic slippage
to electrokinetics}

\author{Laurent Joly$^{1}$} 
\email{ljoly@lpmcn.univ-lyon1.fr}
\author{Christophe Ybert$^{1}$} 
\author{Emmanuel Trizac$^{2}$} 
\author{Lyd\'eric Bocquet$^{1}$}

\affiliation{
$^{1}$ Laboratoire P.M.C.N., UMR CNRS 5586, Universit\'e Lyon I, 69622 Villeurbanne, France\\
$^{2}$ L.P.T.M.S, UMR CNRS 8626, B\^atiment 100, Universit\'e Paris XI, 91405 Orsay, France}



\date{\today}

\begin{abstract}
Hydrodynamic behavior at the vicinity of a confining wall is closely
related to the friction properties of the liquid/solid interface.
Here we consider, using Molecular Dynamics simulations, the electric
contribution to friction for charged surfaces, and the induced
modification of the hydrodynamic boundary condition at the 
confining boundary.
The consequences of liquid slippage for electrokinetic phenomena,
through the coupling between hydrodynamics and electrostatics within
the electric double layer, are explored. Strong amplification of
electro-osmotic effects is revealed, and the non-trivial effect of
surface charge is discussed.
This work allows to reconsider existing experimental data, concerning
$\zeta$ potentials of hydrophobic surfaces and suggest the possibility
to generate ``giant'' electro-osmotic and electrophoretic effects,
with direct applications in microfluidics.
\end{abstract}


\maketitle

\section{Introduction}
\label{sec:intro}

With the important development of microfluidic systems,
miniaturization of flow devices has become a real challenge
\cite{stone04}. Microchannels are characterized by a large
surface-to-volume ratio, so that flows are strongly affected by
surface properties. A clear
understanding of liquids dynamics close to solid surfaces
is consequently an important prerequisite for further progress.
Over the recent years,  important advances 
in  the rheology of fluids at small scales have been performed, partly thanks to computer
simulations, such as Molecular Dynamics (see \eg \cite{barrat99b} and
refs. therein), but mainly thanks to the development of new
experimental techniques, such as optical velocimetry (see
\cite{joseph05,schmatko05} and refs. therein), or dissipation
measurements using Surface Force Apparatus and Atomic Force
Microscope (see \cite{cecile05,vino06} and refs. therein).

In this context, the usual assumption of a no-slip boundary condition
for simple liquids at solid surfaces has been critically revisited at
small scales in the last years, see e.g. \cite{lauga05} for a review.
The conclusions emerging from these studies are that, while the
continuum hydrodynamics theory surprisingly remains valid \emph{up to
  very small length scales}, the \emph{no-slip} boundary condition 
(BC) for the fluid velocity at the solid surface may be violated in
many situations (see e.g.
\cite{cecile05,lettre_exp,barrat99b,granick03,vino06}). 
Moreover, it has been shown that this violation of the
usual no-slip BC is controlled by the wetting properties of the fluid
on the solid surface: while the no-slip BC is fulfilled on hydrophilic
surfaces, a finite velocity slip is measured on hydrophobic surfaces
\cite{barrat99b,cecile05,lettre_exp}, originating in a low friction
of the liquid at the wall.

In this work, we consider the role of electric properties on liquid-solid friction,
a point which has been barely explored up to now \cite{kim05,kim06}. 
Surfaces indeed usually release charges
when in contact with a polar solvent such as  water, which in turns
strongly  modifies the liquid-solid interactions at the interface.
The natural length scale characterizing the electric interaction range in
electrolytes is the so-called Debye length.
This length being typically nanometric in standard
aqueous electrolytes, one can anticipate that
the dynamics of charged systems should
probe hydrodynamics in the nanometric vicinity of charged solid
surfaces. One can in particular expect an interesting coupling
with nanometric slippage, as predicted
theoretically for neutral surfaces \cite{barrat99b} 
and evidenced experimentally \cite{cecile05,lettre_exp}.
Furthermore, hydrophilic and hydrophobic surfaces
exhibit different electric properties \cite{netz04}, and the
coupling between pure wetting effects and ``charge-mediated''
effects is {\it a priori} subtle and remains to be clarified.
Eventually, such an interplay is expected to affect interfacial transport
of charges, {\it i.e.} electrokinetics, which is commonly used to
manipulate liquids in microsystems (\eg electrophoresis and
electro-osmosis). These different points will be considered in the
present paper. To study the relevant length scales involved,  
extensive molecular dynamics (MD)
simulations have been used.

In a previous article \cite{lettre_ek}, we presented first
results concerning the influence of {surface hydrodynamic properties (as encompassed 
in the so-called hydrodynamic boundary condition)}
on electrokinetic effects, focusing on streaming current
experiments and with a restricted set of electric parameters. 
{Beyond the generalization to other electrokinetic effects, the purpose of the present work is to extend this previous analysis by extensively exploring the influence of the various electric parameters on both the static and the dynamic properties of the surface, therefore rationalizing the interplay between surface charge, hydrodynamics at the interface and electrokinetics response.}  
The paper is organized as follows. 
In section \ref{sec:model}, we describe our numerical model, 
together with
some details of the simulation procedure. 
We focus in section \ref{sec:es} on the static properties
of our systems. We then turn to the dynamic behaviour and explore 
in section \ref{sec:dynamique}
the coupling  between liquid/solid friction and surface
charge. The reciprocal coupling
between hydrodynamics and electric properties of the interface
is studied in section \ref{sec:ek}. 
A strong amplification of electrokinetic effects in the presence of
slip at the solid surface is reported. Finally experimental
consequences of this work are addressed 
in section \ref{sec:exp}, with focus on 
the origin of $\zeta$ potential on hydrophobic
surfaces, and the
possibility to strongly amplify electrokinetics effects using
polarized hydrophobic surfaces, with direct applications in
microfluidics.

\section{Model and Parameters}
\label{sec:model}

\begin{figure}
\includegraphics[width=7cm]{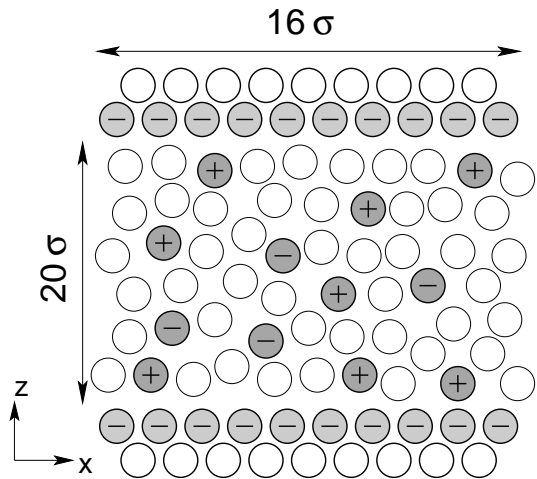}
\caption{Simulated system. The simulation cell extends over
$\units{16}{\sigma}$ along the $\mathrm{O}y$
direction, perpendicular to the figure.\label{fig:config}}
\end{figure}

We first describe our microscopic model (see Fig. \ref{fig:config}) 
and some details of the
simulation procedure.
The fluid system  (solvent and microions) is
confined between two parallel solid substrates, themselves composed of 
individual atoms fixed on an fcc lattice. The solvent and solid
substrate particles interact via Lennard-Jones (LJ) potentials, 
\begin{equation}
v_{ij}(r)=4 \epsilon \left[ \left(\frac{\sigma}{r}\right)^{12} - c_{ij}  
\left(\frac{\sigma}{r}\right)^{6} \right] ,
\label{LJ}
\end{equation}
with identical interaction energies $\epsilon$ and molecular diameters
$\sigma$. The tunable parameters $c_{ij}$ enable us to adjust the wetting
properties of the fluid on the substrate \cite{barrat99b}: for a given
fluid-fluid cohesion $c_{FF}$, the substrate displays a
``hydrophilic'' behavior for large fluid-solid cohesivity, $c_{FS}$,
and a ``hydrophobic'' behavior for small $c_{FS}$.
Here, the wetting (respectively non-wetting) situation is typically
achieved by taking $c_{FS}=1$ (resp. 0.5)  for a fixed
$c_{FF}=1.2$. This leads to a contact angle $\theta$ of a liquid
droplet on the substrate, measured in the simulations, equal to
$80^\circ$ (resp. $140^\circ$) for a temperature $\kt /\epsilon=1$
(see \cite{barrat99b} for an exhaustive discussion on this
point).

Besides, microions interact through both Lennard-Jones potentials
as described in Eq. (\ref{LJ}) and Coulomb potential in a medium with
dielectric permittivity $\ed$:
\begin{equation}
v_{\alpha\beta}(r)= \kt \, q_\alpha q_\beta \, \frac{\lb}{r},
\end{equation} 
where $q_\alpha$ and $q_\beta$ are the valences of the
interacting charges, and  $\lb= e^2/(4\pi \ed \kt)$ is the Bjerrum
length, $e$ denoting the elementary charge (the Bjerrum length is the
typical scale at which thermal energy and electric interaction energy
compare; in water at room temperature $\lb \approx \units{0.7}{nm}$). 
The influence of the solvent permittivity on static and
dynamical properties will be tested by varying $\lb$ in the parameter
range $[0.25 \sigma; 5 \sigma]$; otherwise we will choose $\lb =
\sigma$ as the default value.
Note that we consider a case where LJ parameters are identical for
charged and neutral atoms (liquid or solid); as we verified in
appendix \ref{sec:wetting}, this simplifying assumption enables us to
clearly separate hydrodynamic and electrostatic properties, without
affecting the generic mechanisms evidenced in this work.

Wall atoms are organized into five layers of an fcc solid (in the 100
direction) in both walls. For each wall, only the first layer, which is in
contact with the fluid, is charged. The corresponding $N_\text{wall}$
atoms bear a discrete charge, with
valency $q_\text{wall}=-Z/N_\text{wall}$ so that each wall bears a
negative net charge $-Z e$.
The solvent contains $2Z$ monovalent counterions, to which
$N_\text{s}=N_++N_-$ salt ions are added, all with unit valence. Global
electroneutrality is enforced by imposing $N_+=N_-$. The simulated
systems are generally made up of $10^4$ atoms.
A typical solvent density is $\rhof \sigma^3\sim 0.9$, while the
concentration of microions $\rhos = N_\pm/\mathcal{V}$ will be varied
between $\rhos \sigma^3= 5 \times 10^{-3}$ and $\rhos \sigma^3= 0.16$ (with
$\mathcal{V}$ the total volume of the sample).
With a typical value $\sigma=\units{0.5}{nm}$, this corresponds
roughly 
to an ionic strength varying between $\units{10^{-2}}{M}$
and $\units{1}{M}$. For $\lb = \sigma$, the corresponding Debye
screening length (see 
below) ranges from a few $\lb$ to a fraction of $\lb$.
Salt-free situations have also been investigated and will be
reported below.
The influence of surface charge will be considered, by varying the
charge per unit surface $-\varSigma$ in the parameter range
$[-0.02e/\sigma^2;-0.8e/\sigma^2]$ (note that for convenience we define
the parameter $\varSigma$ to be the negative of the surface charge).
Unless otherwise stated, we will choose $\varSigma = 0.2 e/\sigma^2$, 
with a corresponding Gouy-Chapman length $\lgc = 1/(2\pi \lb
|\varSigma|) = 0.8\sigma$. For $\sigma=\units{0.5}{nm}$, this translates into a typical 
surface density $\units{-0.13}{C/m^2}$. 
For
$\lb=\sigma$ and the salt concentrations considered here, 
the surface potential
$V_0$ ranges between $e V_0 \approx \kt$ and $e V_0
\approx 4 \kt$, allowing us to explore both linear 
(Debye-H\"uckel-like) and non-linear situations.
Periodic boundary conditions are applied in the $x$ and $y$ directions
with $L_x=L_y=16 \sigma$, and the distance between the walls is $L_z=
20.9 \sigma$. Ewald sums are used to compute Coulombic interactions
(assuming a periodicity in the $z$ direction with a box size of $5
L_z$) \footnote{We have used the MD code \textsc{lammps} 2001, by
  S. J. Plimpton\,\cite{lammps}, available at
  http://www.cs.sandia.gov/$\sim$sjplimp/lammps.html.}. 
  In the subsequent analysis, Lennard-Jones units are used, 
  with a characteristic distance
  $\sigma$ and time $\tau=(m\sigma^2/\epsilon)^{1/2}$.
Temperature is kept constant to $\kt = 1$ by applying a Hoover drag
to the $y$ degrees of freedom only, \ie in the direction perpendicular
to the flow and confinement \cite{barrat99b}.

Our model therefore includes the discrete nature of the solvent and
charges and a tuning wettability of the surface, whereas these effects
are usually neglected in the traditional description of electrokinetic
phenomena. We chose to describe Coulombic interactions at
the level of an effective dielectric medium (with dielectric
permittivity $\ed$). This simplifying assumption -- which could be
relaxed using a more realistic model for the solvent \cite{qiao04} --
enables us to investigate specifically the generic interplay between {hydrodynamic}
and electrostatic effects, which is the main focus of this
work. We expect the conclusions obtained to be generically valid.

In order to model the contribution of electric charges to friction, we
will explore the influence of various electric parameters:
electrolyte concentration $\rhos$, surface charge
$\varSigma$ and permittivity of the solvent, \via the Bjerrum length
$\lb$. To be specific, we performed different sets of simulations, including:
\begin{itemize}
\item varying salt concentration ($\rhos \sigma^3 \in [5 \times
  10^{-3};0.16]$), with fixed Bjerrum length ($\lb=\sigma$) and
  surface charge ($\varSigma=0.2e/\sigma^2$);
\item varying surface charge ($\varSigma \sigma^2/e \in
  [0.02;0.8]$), with fixed Bjerrum length
  ($\lb=\sigma$) and salt concentration ($\rhos \sigma^3=0.06$);
\item varying Bjerrum length ($\lb/\sigma \in [0.25;5]$), with fixed surface
  charge ($\varSigma=0.2e/\sigma^2$) and \emph{no electrolyte}. In
  addition, we finally considered the effect of doubling the Bjerrum
  length ($\lb=2 \sigma$) in the presence of salt, with $\rhos
  \sigma^3=0.06$ and $\varSigma=0.2e/\sigma^2$.
\end{itemize}

\section{Static}
\label{sec:es}

We now turn to the results of the simulations. We first focus on the
equilibrium properties of the charged interfaces.
As a rule, a solid surface immersed in an electrolyte solution develops
spontaneously an electric charge, under the action of several
mechanisms: dissociation of ionisable groups, release of ionic
impurities, specific adsorption of charged species present in the
solution, etc.
In response to this surface charge, the microions of the liquid rearrange themselves to form, in the vicinity 
of the solid surface, a diffuse layer named Electric Double Layer (EDL), carrying a net charge opposite to that of the surface.
The so-called Debye length, denoted here $\ld$,
provides a measure of the EDL extension and
determines the electric interaction range between macromolecules. 
This characteristic length plays a crucial role in 
 the static phase behavior of these systems
\cite{lyklema}.

The standard description of the microions cloud in the vicinity of a
charged surface, in the framework of the Gouy-Chapman theory,
involves a Poisson description of the electrostatics ($\Delta V +
\rhoe/\ed = 0$, with $V$ the electric potential and
$\rhoe=e(\rho_+-\rho_-)$ the charge density, defined in terms of the
microions concentrations), coupled with a Boltzmann equilibrium
description for the microions distribution ($\rho_\pm =
\rhos^\text{bulk} \exp (\mp \beta e V)$, with $\beta = 1/\kt$ and
$\rhos^\text{bulk}$ the bulk microions concentration), leading to the
standard Poisson-Boltzmann (PB) equation for the electric potential
$V$ in the EDL \cite{andelman99}:
\begin{equation}
\beta e \Delta V = \kappa^2 \sinh (\beta e V),
\end{equation}
where $\ld = (8\pi\lb\rhos^\text{bulk})^{-1/2}$ is the Debye screening
length. In situations of extreme confinement where the notion
of bulk becomes irrelevant, $\rhos^\text{bulk}$ should be viewed as
a normalization density, or equivalently as the salt density 
in a salt reservoir against which the solution is dialyzed.

\begin{figure}
\includegraphics[width=7cm]{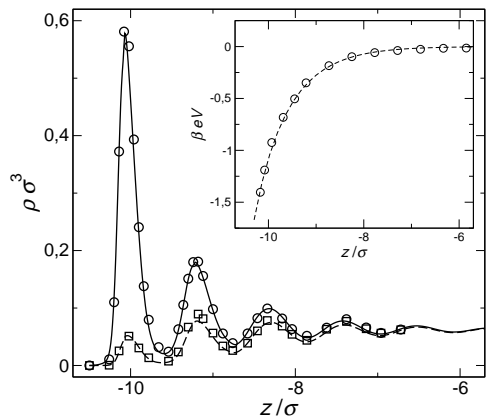}
\caption{Microionic density profiles, averaged over the $xy$
  directions ($\rhos\sigma^3=0.06$, $\varSigma=0.2e/\sigma^2$,
  $\lb=\sigma$, wetting case). Symbols: Molecular Dynamics results 
  for the counter-ions ($\circ$) and co-ions
  ($\scriptscriptstyle\square$); solid and dashed lines correspond to the
  predictions of the modified PB description (see text). The inset shows the
  electrostatic potential. Symbols ($\circ$): Molecular Dynamics results calculated
  from Poisson's equation and the measured microions profiles; dashed line: bare
  PB prediction (see text). The position of the wall, defined as that
  of the centers of the last layer of wall atoms, is located at
  $z_\text{wall}= -10.9 \sigma$.\label{fig:es}}
\end{figure}

In Fig. \ref{fig:es}, we show typical
density profiles of the microions close to one of the confining
surfaces.
Important structuration effects can be observed close
to the charged surface. This is {\it a priori} incompatible with the
Poisson-Boltzmann (PB) prediction \cite{andelman99}: within PB, 
an oscillating density profile implies that the electric field vanishes at 
every extrema, which corresponds to a change of sign of the integrated
charge. This in turns implies that the system exhibits overcharging.
Such an effect can be rigorously ruled out within mean-field descriptions,
such as the PB one \cite{pre2000}. 
 However, the oscillations in the microions profiles exhibited here
 are not associated with any charge inversion, but 
result from the structuration
of the solvent itself.  Such an effect can be captured by a
modified PB description. Indeed, due to the presence of the solvent
particles, the microions not only organize themselves due to electrostatic
interactions (which corresponds to the usual PB description) but also
due to the effective external field associated with the structuration
in the solvent, $V_\text{ext}(z)=-\kt
\log\left[\rhof(z)/\rhof\right]$, with $\rhof(z)$ the solvent density
profile and $\rhof$ its bulk value.
The microions density profiles $\rho_\pm(z)$ correspondingly
obey a modified Boltzmann equilibrium:
\begin{equation}
\rho_\pm(z) \propto e^{\beta(\mp  e V(z) - V_\text{ext}(z)) }\propto
\rhof(z) e^{\mp \beta e V(z)} .
\label{rhopm}
\end{equation}
Such a relationship emerges naturally from a
simple Density Functional Theory, that allows to rationalize the
argument, accounting for the the discrete nature
of both solvent and charged atoms exactly,
while the standard mean-field PB free energy is assumed for the
electrostatic part. The details can be found in appendix \ref{app:C}.

Inserting Eq. (\ref{rhopm}) into Poisson equation, we find that the
electrostatic potential follows a modified PB equation, 
\begin{equation}
\beta e \Delta V = \kappa^2 \gamma(z) \sinh(\beta e V)
\label{PB_mod}
\end{equation}
where 
$\gamma(z)=\rhof(z)/\rhof$ is the normalized {\it solvent} density
profile. This equation allows for instance to compute the electric potential
once the solvent density is known, but should be supplemented
with a closure relation to {\em predict} both $\rho_f(z)$ and $V(z)$.
A more modest goal is to test the relevance of this approach by 
measuring the fluid density profiles, $\rhof(z)$, and subsequently 
solving Poisson
equation with the microionic densities given by Eq. \eqref{rhopm}. As
shown in Fig. \ref{fig:es}, this procedure leads to results that 
are in remarkable agreement with simulations profiles obtained from Molecular
Dynamics. Moreover, a further useful
approximation can be proposed : the solution of the modified PB
equation for the electrostatic potential is as a matter of fact very
close by the ``bare'' PB solution $V_\text{PB}(z)$
(corresponding to $\gamma(z)=1$), whose analytic expression can be
found in the literature \cite{lyklema,andelman99,hunter_zeta}. 
This leads to $\rho_\pm(z)
\propto \rhof(z) \exp[\mp \beta e V_\text{PB}(z)]$. The validity of
this approximation -- surprising in view of the strong layering effect
at work -- is emphasized in Fig. \ref{fig:es} (inset), where the
corresponding bare PB potential \cite{hunter_zeta} is plotted against
the ``exact'' electrostatic potential. The latter is obtained from the
simulations using Poisson's equation by integrating twice the charge
density profile $\rhoe = e(\rho_+-\rho_-)$.

\begin{figure}
\includegraphics[width=7cm]{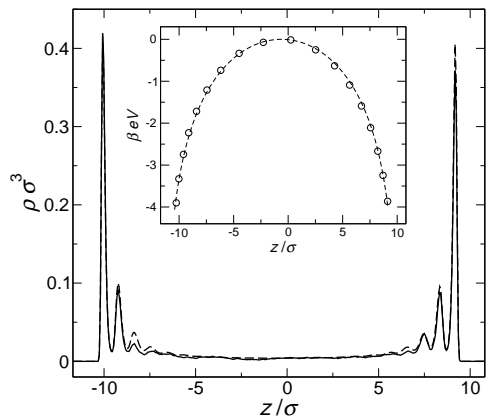}
\caption{Counterions density profile, averaged over the $xy$
  directions, without added  electrolyte ($\Sigma = 0.2 e/\sigma^2$,
  $\lb = \sigma$, wetting case). (---): MD results; ($--$): prediction
  of the modified PB description, using the bare PB potential (see
  text). Inset: electrostatic potential. Symbols ($\circ$): MD results
  calculated from Poisson's equation and counterions profiles; dashed
  line: bare PB prediction (see text). The walls are located at
  $z_\text{wall}^\text{inf} = -10.9 \sigma$ and
  $z_\text{wall}^\text{sup} = 10.0 \sigma$.\label{fig:es_sans_sel}}
\end{figure}

The case of no-added salt gives another interesting limiting case, which
we now consider.
In the no-salt case, only counterions are present in the solution,
with a total charge compensating exactly the charge of both
surfaces. The PB equation becomes:
\begin{equation}
\beta e \Delta V = -4\pi \lb e \rho_0 \text{e}^{-\beta e V} ,
\end{equation}
where $\rho_0$ represents a reference concentration for counterions,
for which $V=0$. This equation can be solved analytically for the
electric potential $V_\text{PB}$ and the counterions concentration
$\rhoi$ in the simple case of an electrolyte confined between two
parallel plates, as considered in our simulations
\cite{andelman99}. As in the presence of salt, we can extend the PB
prediction to take into account the effective external field due to
the structuration of the solvent, using the very same approach. The
results for a typical no-salt configuration are presented on
Fig. \ref{fig:es_sans_sel}. 
We can check that the bare PB prediction
perfectly accounts for the electric potential, for which no
structuration can be seen. The counterions density profile is
accordingly well described by $\rhoi(z) = \rhof(z) \exp [-\beta e
  V_\text{PB}(z)]$.

 Globally we have tested the validity of this modified PB approach 
 in our simulations with various conditions, involving a broad range 
 of the different parameters,
salt concentration $\rhos$, surface charge
$\varSigma$ and Bjerrum length $\lb$.
The modified PB approach has been found to describe with accuracy every
situation considered, over the whole parameter range
for both surface charge and salt concentration (not shown).
However, we found that the modified PB prediction breaks down at
large Bjerrum length (we considered two cases with 
$\lb=2.24\sigma$ and $\lb=5.04\sigma$, not shown). 
This failure is however expected : for large Bjerrum length, 
microionic correlations become important
and invalidate the modified Poisson-Boltzmann ansatz. A simple
analysis, discussed in details
at the end of appendix \ref{app:C}, shows that the criterion for 
this failure can be written $(\varSigma/e)\lb^{\,2} > 1$.

\section{Dynamics}
\label{sec:dynamique}

\subsection{Wetting versus non-wetting : the Poiseuille test-bench}
We now come to the dynamical aspects.
As a first step, we briefly recall how the {hydrodynamics at the interface, as characterized by the }hydrodynamic boundary condition (HBC), is affected by a
modification of the wetting properties of the surfaces (see
\cite{barrat99b,barrat99a,Bocquet94} for a more detailed discussion);
We then turn to the specific role of electric parameters.

\begin{figure}
\includegraphics[width=7cm]{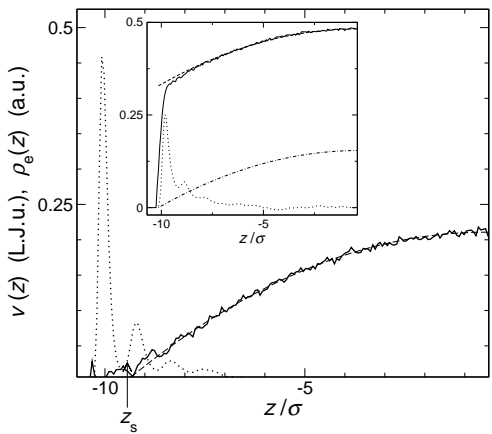}
\caption{Measured Poiseuille velocity profile (solid line) in the
  wetting case ($c_{FS}=1$, $\lb=\sigma$, $\varSigma=0.2e/\sigma^2$,
  $\rhos\sigma^3=0.06$). Dashed line: hydrodynamic prediction 
  using a no-slip BC at the 'plane of shear' located at $z_\text{s}$
  (indicated by the arrow). To emphasize the existence of an immobile
  Stern layer, we also indicate the charge density profile
  $\rhoe(z)=e(\rho_+(z)-\rho_-(z))$ (dotted line), with arbitrary
  units. The position of the wall (defined as that of the centers of
  the last layer of wall atoms) is at $z_\text{wall} = -10.9
  \sigma$. Inset: 
  Results for the non-wetting case ($c_{FS}=0.5$). Solid line:
  velocity profile measured in the simulation (shown on the same scale
  as in the main graph); dashed line: hydrodynamic prediction with a
  partial slip BC, with a slip length $b \approx 11 \sigma$; dashed-dot line:
  hydrodynamic prediction with a no-slip BC; dotted line: charge
  density profile (arbitrary units).\label{fig:dynamique}}
\end{figure}

We chose to probe the HBC in a Poiseuille configuration, by applying
an external force per particle $f_0$, in the $x$ direction, to all
microscopic particles. The Bjerrum length and surface charge are set to
their default values ($\lb=\sigma$ and $\varSigma=0.2e/\sigma^2$), and
the salt concentration is varied between $\rhos \sigma^3 = 5\times
10^{-3}$ and $\rhos \sigma^3 = 0.16$.
We start by discussing the measured velocity profiles. The situation
corresponding to a wetting substrate -- with $c_{FS}=1$ in
Eq. \eqref{LJ} -- is shown in the main plot of Fig. \ref{fig:dynamique} 
(here for $f_0=0.02$ in LJ units).
The velocity profile shows a parabolic shape as
predicted by continuum hydrodynamics, even at the EDL level.
Moreover, the viscosity, deduced from the curvature of the
parabolic shape, retains its bulk value. Nevertheless, our
measurements have shown that the
no-slip BC applies inside the liquid, {\it at a distance of
  about one layer of solvent particle}. This observation is consistent
with previous theoretical {and experimental} predictions \cite{Bocquet94,barrat99b}. This
position of the no-slip BC here defines the ``plane of shear''
position, $z_\text{s}$, usually introduced in the electrokinetic
literature \cite{hunter_zeta}. We note that $z_\text{s}$ does not vary
significantly with the salt concentration, in the parameter range 
investigated.


On the other hand, the non-wetting case displays a very different
behavior, as shown in the inset of Fig. \ref{fig:dynamique}. A non-wetting
substrate is set up by choosing a smaller value of $c_{FS}$, here
$c_{FS}=0.5$. First, concerning the velocity profile, a large amount
of slip is found at the wall surface, in accordance with observations
on non-wetting surfaces \cite{barrat99b}. More quantitatively,
slippage is characterized by a {\it slip length}, $b$, defined as the
distance at which the linear extrapolation of the velocity profile
vanishes. In other words, this amounts to replace the no-slip BC by a
partial slip BC, defined as $b\,\partial_z v = v$ at the wall position
\cite{Bocquet94}. As shown in the inset of Fig. \ref{fig:dynamique}, the
velocity profile is well fitted by the continuum
hydrodynamics (parabolic) prediction, together with a partial slip BC,
characterized by a non vanishing slip length (here $b \approx 11
\sigma$). The measured slip length $b$ barely depends on
the salt concentration, a point which we now rationalize.

\subsection{Rationalizing friction and slipping}
\label{ssec:ratio}

Slippage can be interpreted in terms of friction properties at the liquid/solid interface.
Indeed, the usual partial slip BC can be interpreted as the
continuity of tangential stress at the liquid/solid interface: 
the viscous shear stress exerted by the liquid on the wall $\eta\,
\partial_z v$ (where $\eta$ is the liquid viscosity) is equal
to the friction stress suffered by the liquid from the wall,
which can be written in the form $\sigma_{xz} = \lambda v_\text{s}$, where
$\lambda$ is the interfacial friction coefficient, linking the
friction stress $\sigma_{xz}$ and the relative liquid/solid slip
velocity $v_\text{s}$.
This equality corresponds to the partial slip BC $v_\text{s} = b\,
\partial_z v$, with the slip length $b = \eta / \lambda$. This
expression formalizes the simple idea that liquids slip all the more 
as the interfacial friction is low. It also means that the slip length
characteristics can be directly deduced from a friction analysis at
the liquid/solid interface.

The dependence of the slip length on the electric parameters can
then be rationalized on the basis of
a simple argument based on the influence of the electric interaction
on the friction coefficient.
We start from a Green-Kubo expression for the interfacial friction
coefficient\,\cite{Bocquet94}:
\begin{equation}
\lambda = \frac{\eta}{b} = \frac{1}{\mathcal{A}\,\kt} \int_0^\infty
\langle F_x(t)F_x(0) \rangle \mathrm{d}t,
\label{lambda_green_kubo}
\end{equation}
where $\mathcal{A} = L_x L_y$ is the area of the solid surface under
consideration, and $F_x$ is the $Ox$ component of the instantaneous
force exerted by the wall on the liquid at equilibrium. Apart from a
few simple situations, it is difficult to evaluate this
expression. In this study we therefore restrict ourselves to
extracting scaling laws for the slip length $b$.
We start with the evaluation of an order of magnitude for the temporal
auto-correlation integral:
\begin{equation}
\int_0^\infty \langle F_x(t)F_x(0) \rangle \mathrm{d}t =
F^2_x \tau_D  ,
\label{integrale}
\end{equation}
where $F_x = \sqrt{\langle F_x^2(t) \rangle}$ is the r.m.s. force, 
and $\tau_D$ the relaxation time scale of the force correlation
function. The latter can be estimated as the diffusion time
of the liquid molecules over the characteristic wavelength
of the wall corrugation $\ell_\text{r}$ \cite{barrat99a}:  
\begin{equation}
\tau_D= {\ell^2_\text{r} \over D}
\end{equation}
where  $D$ is
the self-diffusion coefficient of liquid molecules. The validity
of this estimate has been exhaustively tested in Ref. \cite{barrat99a}.

We now separate Lennard-Jones and
electric contributions to the total force: $F_x =
F_\text{LJ}+F_\text{ES}$. Using Eq. \eqref{lambda_green_kubo}, we then
write: $1/b \propto (F_\text{LJ}+F_\text{ES})^2 \propto
F^2_\text{LJ}+F_\text{LJ} F_\text{ES}+F^2_\text{ES}$.
In every performed simulations, we noticed that the slip properties
were only slightly affected by the presence of charge; we can therefore assume
that the electric contribution to friction is small compared to the
Lennard-Jones term; it is then possible to neglect the pure
electric contribution (the $F^2_\text{ES}$ term), which leads to:
\begin{equation}
\frac{1}{b} = \frac{1}{b_\text{LJ}} + \frac{1}{b^\prime} ,
\label{eqn:decomp_b}
\end{equation}
with $1/b^\prime \propto F_\text{LJ} F_\text{ES}$.
We can then estimate an order of magnitude for the first order electric
contribution to friction: $1/b^\prime \propto F_\text{ES} \sim Q \times E$,
where $Q=\varSigma \mathcal{A}$ is the total charge of the EDL
(compensating the surface charge), and $E=\varSigma/\ed\sim \varSigma
\lb$ the electric field at the interface. We finally find that the
$b^\prime$ contribution to slippage varies as:
\begin{equation}
b^\prime \sim \varSigma^{-2} \lb^{-1} ,
\label{eqn:b_prime}
\end{equation}
which we rewrite $b'/\sigma = \alpha (\varSigma \sigma^2/e)^{-2}
(\lb/\sigma)^{-1}$, $\alpha\sim 1$ a numerical prefactor.
The electric contribution to the slip length therefore does not depend on the Debye 
length of the system.
Alltogether, we obtain the following prediction for the slip length $b$ on the
surface charge~:
\begin{equation}
b= {b_{LJ}æ\over {1+{\varSigma^{2} \over e^2} \lb \sigma^2 b_{LJ} }}
\label{eqn:b_full}
\end{equation}
Note that the dependence of the Lennard-Jones contribution to the
slip length, $b_{LJ}$, on
the microscopic parameters can also be predicted, see Ref. \cite{barrat99a}.

\subsection{Comparison of simulation data against theory for slip
length and shear plane position}

We have tested the above description against MD results.  To this end,
we extracted the first order electric
contribution $b'$ from the simulation results, using
Eq. \eqref{eqn:decomp_b}; to this purpose, the LJ contribution
was \emph{measured} in complementary simulations, using identical
systems \emph{without charges} (with identical pressure).
\begin{figure}
\includegraphics[width=7cm]{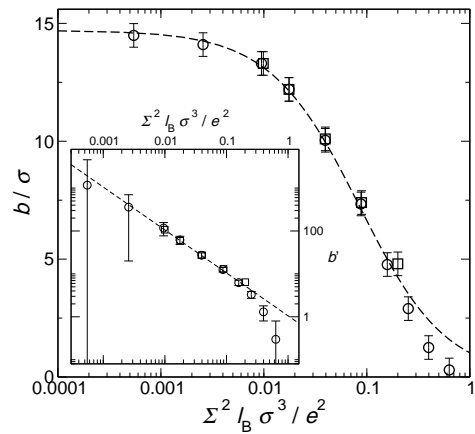}
\caption{Slip length $b$ as a function of $\varSigma^2 \lb$, in the
  non-wetting case. Symbols: MD results with fixed Bjerrum length
  $\lb=\sigma$ and varying surface charge ($\circ$) ($\rhos \sigma^3 =
  0.06$), or fixed surface charge $\varSigma=0.2e/\sigma^2$ and
  varying Bjerrum length, \emph{without added salt}
  ($\scriptscriptstyle\square$). Dashed line: analytical
  fit using Eq. (\ref{eqn:b_full}) (see text).
  Inset: First order electric contribution $b^\prime$ as a
  function of $\varSigma^2 \lb$. Symbols: MD results (symbols are
  identical to those of the main graph). Uncertainties on results with
  varying Bjerrum length, not represented for the sake of clarity, are
  comparable to the size of the symbols. The dashed line has a slope $-1$. 
  The agreement is excellent except for very high
  surface charge (see text). 
  \label{fig:b_charge}}
\end{figure}
The inset of Fig. \ref{fig:b_charge} presents the numerical results
for $b'$ as a function of $\varSigma^2 \lb$, using various $\varSigma$
and $\lb$, which are perfectly fitted with the prediction of our
simple model: $b'/\sigma = 1.10 (\varSigma \sigma^2/e)^{-2}
(\lb/\sigma)^{-1}$. Only with the highest surface charges does a slight
discrepancy appear 
\footnote{A possible reason is that for such high charges, the pure electric contribution 
$F_{\text{ES}}^2$ cannot be neglected which leads to a decrease of $b$, 
as observed in Fig. \ref{fig:b_charge}.}. 
We were thus able to account precisely for the complete slip length $b$
with an analytical expression, Eq. (\ref{eqn:b_full}), simply by adding the LJ and electric
contributions, as can be seen on the main graph of
Fig. \ref{fig:b_charge}. Finally we have checked that the slip length is 
independent on the salt concentration at fixed $\varSigma$ and
$\lb$ (not shown), again in accordance with our simple model.

In the wetting case, no slip occurs at the surfaces; yet we can
consider the position of the no-slip plane $z_\text{s}$, defining
equivalently the width of the immobile Stern layer (see section
\ref{sec:ek}). 
\begin{figure}
\includegraphics[width=7cm]{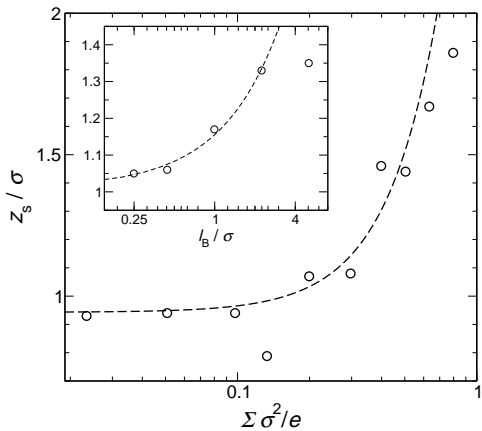}
\caption{Width of the immobile Stern layer $z_\text{s}$ as a function
  of surface charge $\varSigma$ (wetting case, $\lb = \sigma$, $\rhos
  \sigma^3 = 0.06$). Symbols: MD results; dashed line: analytical fit
  with a simple quadratic expression (see text).
  Inset: $z_\text{s}$ as a function of Bjerrum length $\lb$ (wetting
  case, \emph{no salt}, $\varSigma = 0.2 e / \sigma^2$). Symbols: MD
  results; dashed line: simple analytical fit.\label{fig:zs_charge}}
\end{figure}
Figure 
\ref{fig:zs_charge} shows the evolution of $z_\text{s}$ with
$\varSigma$ (main graph) and $\lb$ (inset). {Using a Green-Kubo expression similar to the one used for $b$ \cite{Bocquet94} and following the same derivation as above, one expects $z_\text{s}\sim\varSigma^2/\lb$. In spite of the rather important measurement noise, this prediction is rather well verified as can be seen from figure \ref{fig:zs_charge} where $z_\text{s}
(\varSigma)$ is fitted by $z_\text{s}/\sigma =
0.94 + 2.27 (\varSigma \sigma^2/e)^2$ and $z_\text{s}
(\lb)$ by $z_\text{s}/\sigma = 1.01 + 0.145
\lb/\sigma$. Finally, we recall that no effect of salt concentration on
$z_\text{s}$ has been observed.}


\section{Electrokinetics}
\label{sec:ek}

In section \ref{sec:dynamique}, we have shown that the traditional
no-slip BC for the liquid velocity at the wall was violated on
non-wetting surfaces, with a small but finite slip length ($b \sim 10
\sigma$, corresponding to $b \sim \units{5}{nm}$), even for highly
charged surfaces (significant slip until $\varSigma \sim
-0.3e/\sigma^2$, corresponding to $\varSigma \sim \units{0.2}{C/m^2}$
and $V_0 \sim \units{80}{mV}$).
We now investigate the consequences of these modifications of
hydrodynamics at the interfaces for the dynamics of charged systems,
and particularly for electrokinetic phenomena, commonly used to
manipulate liquids in microsystems (\eg electrophoresis and
electro-osmosis).
After a brief recall of electrokinetic effects, we will perform
streaming current and electro-osmosis simulations, in order to explore
the influence of such hydrodynamic slippage. Finally, we will focus on the
specific role of electric parameters.

\subsection{Zeta potential}

In addition to its interest for the understanding of static properties
of charged systems, the EDL is, on the dynamical level, at the origin
of numerous electrokinetic phenomena: electrophoresis,
electro-osmosis, streaming current or potential, etc. Because these
various effects take their origin at the \textit{surface} of the sample \via
the EDL, they provide smart and particularly efficient ways to drive
or manipulate flows in microfluidic devices \cite{stone04,bazant04},
where surface effects become predominant.

The extension of the EDL is typically on the order of a few nanometers
and electrokinetic phenomena therefore probe the {\it nanorheology} of
the solvent+ions system at the charged surface. This can thus raise
some doubts regarding the validity of continuum approaches to describe the
dynamics at such scales. Those doubts seem particularly relevant
concerning the traditional description of the EDL dynamics, which
relies both on the mean-field Poisson-Boltzmann theory of the
microion clouds, but also on continuum hydrodynamics for the flow
fields \cite{hunter_zeta}. These two aspects are embodied in the
so-called {\it zeta potential}, denoted $\zeta$, which is 
\textit{traditionally} defined as the electric potential $V(z_\text{s})$
computed at the surface of shear $z_\text{s}$, where the fluid
velocity {\it vanishes}. This quantity plays a key role in electrokinetic
phenomena \cite{hunter_zeta,netz03,qiao04,churaev02}, since it
quantifies the coupling between flow characteristics in the solvent
(via the mean velocity or applied pressure drop) and electric
quantities (electric field, induced streaming current or
potential). An important point is that the standard electrokinetic
description is based on the assumption of a no-slip boundary condition
of the liquid at the solid interface.

In section \ref{sec:es}, we showed that the traditional PB
description is relevant for the electric potential, even at the EDL
scale. Yet in section \ref{sec:dynamique}, we observed that {although a continuum hydrodynamic approach stands down to EDL sizes}, the no-slip BC could be violated in non-wetting situations, in accordance
with previous numerical \cite{barrat99b} and experimental
\cite{cecile05,lettre_exp} work. We expect this modification of the
interfacial hydrodynamics to affect electrokinetic properties of
charged interfaces, as it was demonstrated in a previous work
\cite{lettre_ek}, focused on streaming current simulations, and
where only a limited set of electric parameters have
been investigated. We will now extend this
work to other electrokinetic phenomena, exploring extensively the
influence of various electric parameters.

\subsection{Streaming current}
\label{ssec:stream}


First of all, we note that the Poiseuille configurations implemented
in the preceding section to probe the HBC corresponds directly to
\emph{streaming current} experiments: the external force $f_0$
accounts for the pressure gradient, and we can measure the electric
current, $I_\text{e}$, associated with the convective motion of the
microions.

The standard EDL description of this electrokinetic effect
predicts a linear relationship between the current and the force, in
the form \cite{hunter_zeta}:
\begin{equation}
I_\text{e}=-\frac{\ed \zeta}{\eta} \mathcal{A} f_0 ,
\label{Ie}
\end{equation}
where $\eta$ is the shear viscosity of the fluid and $\mathcal{A}$ the
fluid slab cross area.
Linear response (in the applied force)
was carefully checked in our simulations. In the following we
use this expression as the {\it definition of the $\zeta$ potential}, in
line with experimental procedures.

In the wetting case, we have seen that the liquid velocity vanishes
inside the liquid, at a distance $z_\text{s}$ of about one layer of
solvent particles (see section \ref{sec:dynamique}).
As shown in Fig. \ref{fig:dynamique}, where the charge density
profile, $\rho_\text{e}(z)$, is plotted against distance, the first
layer of microions, located within $z_\text{s}$, does not contribute
to the convective transport, thereby reducing the global streaming
current. This first layer coincides with the so-called Stern layer of
immobile microions close to the charged surface \cite{hunter_zeta}.

In the non-wetting case, the liquid was shown to slip
significantly at the wall, with a complete disappearance of the
immobile Stern layer (see section \ref{sec:dynamique}).
Concerning microions transport, an important point here is that the
first layer of microions now contributes by a large amount to the
global streaming current, at variance with the wetting case. 
In other words,
the  plane of shear position $z_\text{s}$ is now virtually located
\emph{beyond} the wall, and the Stern layer has completely
disappeared. The remobilization of the Stern layer adds on to the
slippage effect and contributes significantly to the increase of
the electric current measured for hydrophobic surfaces.

\begin{figure}
\includegraphics[width=7cm]{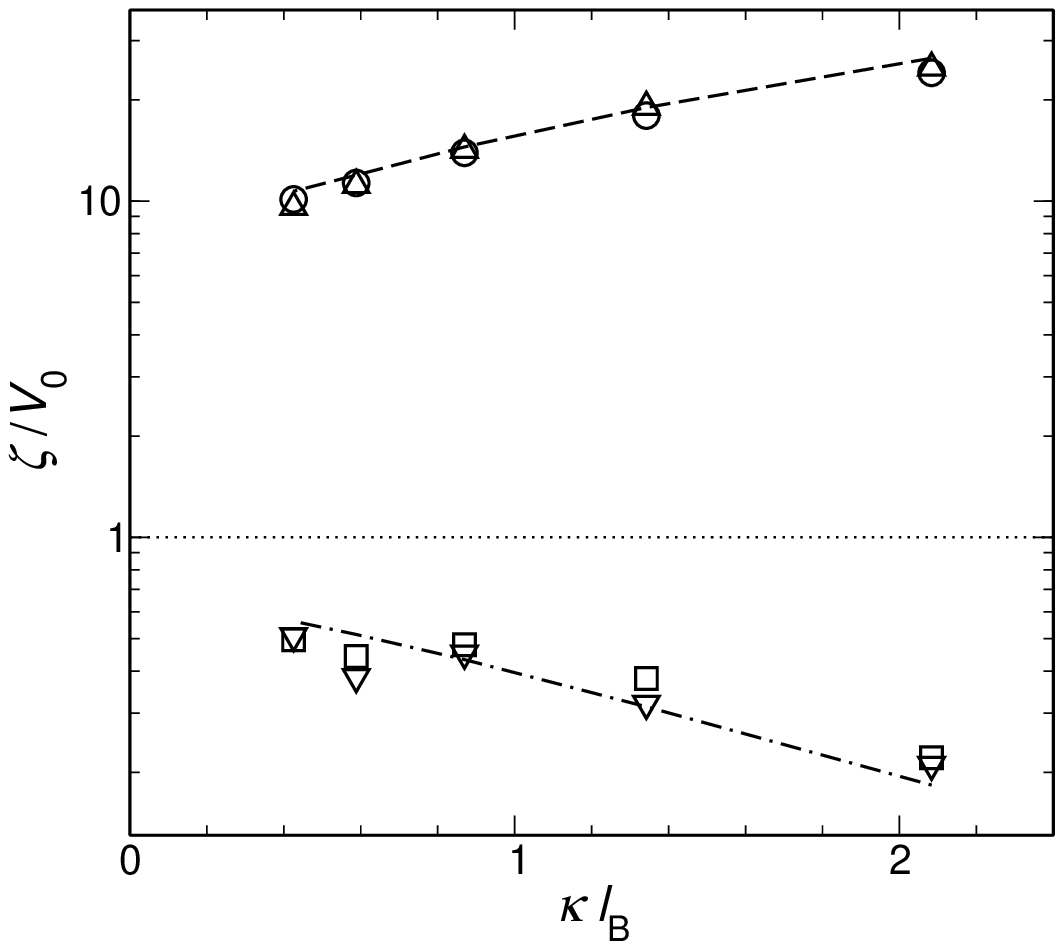}
\caption{Measured $\zeta$ potential as a function of the screening
  factor $\kappa\lb$ in streaming current
  ($\scriptscriptstyle\square$: wetting case; $\circ$: non-wetting
  case) and electro-osmosis ($\scriptscriptstyle\bigtriangledown$:
  wetting case; $\scriptscriptstyle\bigtriangleup$: non-wetting case)
  simulations.
The $\zeta$ potential is normalized by the bare surface potential
$V_0$ obtained from the PB expression at a given $\kappa$ and surface
charge (see the discussion on the inset of
Fig. \ref{fig:es}). For the wetting case (bottom), the dashed-dotted
line is the PB electrostatic potential $V(z_\text{s})$, where the
'plane of shear' position $z_\text{s}$ does not vary significantly
with salt. For the non-wetting case (top), the dashed line
corresponds to the slip prediction Eq. \eqref{eqn:ampliNL}, with $b=11
\sigma$. 
\label{fig:zeta}}
\end{figure}

We summarize our results in Fig. \ref{fig:zeta} and plot the $\zeta$
potential [deduced from the measure of the charge current and
 Eq. (\ref{Ie})] as a function of the Debye screening factor in
the wetting and non-wetting cases. In this figure the $\zeta$
potential is normalized by the bare surface potential $V_0$, 
obtained from the analytic PB expression \cite{hunter_zeta},
as shown \eg in the inset of Fig. \ref{fig:es}.
The overall conclusion from Fig. \ref{fig:zeta} is that non-wettability
strongly amplifies the electrokinetic effects: the ratio between the
$\zeta$ potential and the surface potential is much larger in the
hydrophobic case as compared to the hydrophilic case. More precisely,
in the wetting case the $\zeta$ potential is fixed by the electric
properties of the surface, and coincides with the electric potential
at the ``plane of shear'', $\zeta \approx V(z_\text{s})$, as is usually
assumed \cite{hunter_zeta}. Indeed, 
the simulation points for $\zeta$ are compared to the PB
estimate for the electric potential, $V(z_s)$, showing an overall very
good correspondence.
Conversely, the $\zeta$ potential in the non-wetting case is
dominated by the slip effect and the immobile Stern layer is
completely absent. The effect of {such a modification of the hydrodynamic surface properties} can be accounted for by
considering the partial slip BC in the electrokinetic current
$I_\text{e}=\int \mathrm{d}S \rhoe(z) v(z)$, with $\rhoe(z)$ the charge
density and $v(z)$ the velocity profile, characterized by a slip length
$b$. Within linearized PB description (valid for $eV_0 \ll \kt$),
the result for the current $I_\text{e}$, Eq. \eqref{Ie}, may then be
written $I_\text{e} = (\ed V_0 / \eta) (1 + \kappa b) f_0$
\cite{stone04,churaev02}. For the $\zeta$ potential in the
non-wetting case, this amounts to:
\begin{equation}
\zeta= V_0 (1 + \kappa b),
\label{eqn:ampli}
\end{equation}
with $V_0$ the bare potential of the surface.
A detailed derivation of this effect
is given in Appendix  \ref{sec:zeta_glissant}.

For important potentials ($eV_0 \gtrsim \kt$), a non-linear PB
counterpart of this expression can be obtained (see appendix \ref{sec:zeta_glissant},
section \ref{app:kappaeff}):
\begin{equation}
\zeta= V_0 (1 + \kappa_\text{eff} b),
\label{eqn:ampliNL}
\end{equation}
where the effective Debye length $\kappa^{-1}_\text{eff}$ is
defined by $\kappa_\text{eff} = -\partial_n V (0)/V_0$, and goes to
$\ld$ in the linear limit. 
An equivalent expression has been 
recently discussed in a molecular hydrodynamics study 
of electro-osmosis in clays \cite{dufreche05}.
This expression is
compared to simulation results in Fig. \ref{fig:zeta}, showing again a
very good agreement.

It is possible to retrieve the expression of the amplification ratio
using a simple argument: the streaming current is given
by $I_\text{e} \sim Q \bar{v}$, $Q$ being the charge of the EDL and
$\bar{v}$ the average velocity of the EDL, with
$\bar{v}=\ld \dot{\gamma}$ in the absence of slip,
whereas $\bar{v}=(\ld + b) \dot{\gamma}$ with a slip
length $b$. The amplification ratio is therefore given by
$\bar{v}_\text{slip}/\bar{v}_\text{no-slip} = 1 + \kappa b$.

\subsection{Electro-osmosis}

The streaming current simulations show that electrokinetic
measurements do not probe electrostatic
properties of the system only: when slippage occurs at the walls, the
$\zeta$ potential is much larger that the bare surface potential
$V_0$. This also means that the electrokinetic phenomena used to move
liquids in microfluidic systems could be strongly amplified by
hydrodynamic slippage.
In order to illustrate this interesting outlook, we performed
electro-osmosis simulations, using the same numerical system.
A uniform electric field $E_x$ applied in the channel induces a volume
force inside the EDL, generating \textit{in fine} a plug flow of the
liquid.
The standard description of this phenomenon predicts again a
linear relationship between the electro-osmotic velocity and the applied
field:
\begin{equation}
v_\text{eo} = -(\ed \zeta / \eta) E_x .
\label{eq:eo_standard}
\end{equation}
In the simulation, we imposed an electric force $f_x=qe E_x$ to every
ion ($qe$ being the ion charge), and we measured the resulting velocity
profile in the channel. We then used Eq. \eqref{eq:eo_standard} to
compute the corresponding $\zeta$ potential (the results presented
were obtained for $E_x = 1.0$ in Lennard-Jones units). 
{As for the streaming current, linear response in the applied electric field was checked.}

\begin{figure}
\includegraphics[width=7cm]{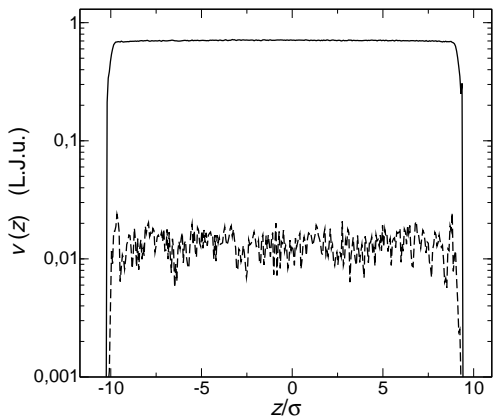}
\caption{Solvent velocity profile $v(z)$, averaged over the $xy$
directions, in a typical electro-osmosis simulation ($\rhos \sigma^3 =
0.06$). An electric field $E_x$ is applied along the $x$
  direction, leading to a plug velocity profile in the
  cell. The dashed line corresponds to the wetting case ($c_{FS}=1$),
  whereas the solid line corresponds to the non-wetting case
  ($c_{FS}=0.5$). The Debye screening factor is $\kappa
  \lb=1.3$.\label{fig:v_eo}}
\end{figure}
Typical velocity profiles are shown in figure
\ref{fig:v_eo}. For both wetting and non-wetting situations, we
observe a plug flow, characteristic of electro-osmosis. Moreover, we 
note
that the electro-osmotic velocity is considerably amplified
-- by almost two orders of magnitude -- in the non-wetting case, all
electric parameters being equal. The $\zeta$ potential, computed
from the measured electro-osmotic velocity using equation
\eqref{eq:eo_standard}, are in perfect agreement with those
obtained in streaming current simulations.

It is possible to recover the slip prediction for $\zeta$ in the
electro-osmosis case by stating that just outside the EDL, the
viscous stress $\sigma_\eta$ compensates the electric
forcing $\sigma_\text{e}$ (integrated over the Debye layer).
Writing $\sigma_\eta = \eta \partial_z v \sim \eta
v_\text{eo}/(\ld+b)$ and $\sigma_\text{e} = Q E_x = -(\ed \kappa V_0) E_x$, $Q =
-\varSigma$ being the net surface charge of the EDL, 
one immediately finds~:
\begin{equation}
v_\text{eo} = -\frac{\ed V_0 (1+\kappa b)}{\eta} E_x = -\frac{\ed
  \zeta}{\eta} E_x ,
\end{equation}
with $\zeta = V_0(1+\kappa b)$. The amplification of the zeta potential
thus originates in the reduction of the velocity gradient in the Debye layer 
by a factor $\ld/(\ld+b)$. 

\subsection{Influence of electric parameters}

\begin{figure}
\includegraphics[width=7cm]{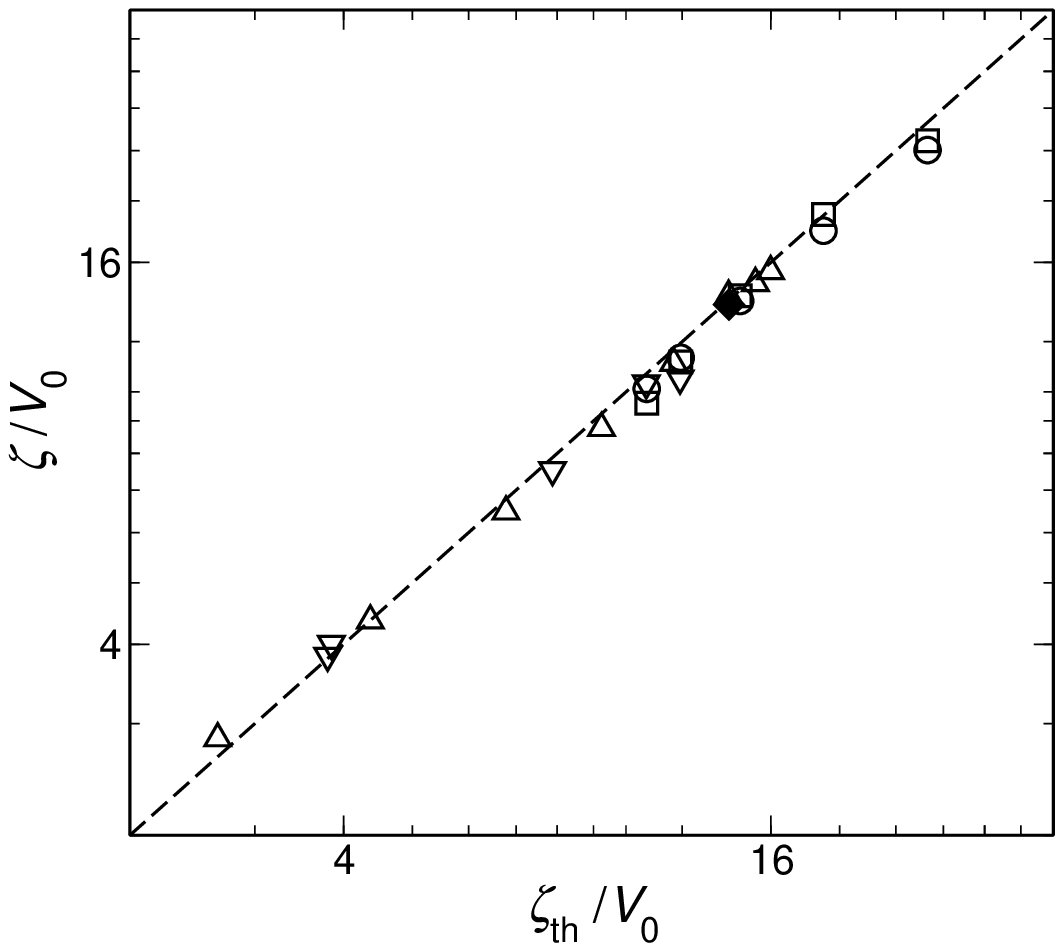}
\caption{Measured $\zeta$ potential versus slip prediction
  $\zeta_\text{th}$ for various situations: previous streaming current
  ($\circ$) and electro-osmosis ($\scriptscriptstyle\square$)
  simulations with fixed $\lb=\sigma$ and $\varSigma = 0.2e/\sigma^2$;
  new streaming current simulations with: fixed $\lb=\sigma$ and
  varying surface charge $\varSigma \sigma^2/e \in [0.1;0.8]$
  ($\scriptscriptstyle\bigtriangleup$); larger Bjerrum length
  $\lb=2\sigma$ ($\scriptstyle\blacklozenge$); non-wetting solvent
  ($c_{FS}=0.5$), various ions wetting properties and concentration
  ($c_{IS} \in [0.6;1.0]$ and $\rhos \sigma^3 \in [0.004;0.14]$)
  ($\scriptscriptstyle\bigtriangledown$). $\zeta$ potentials are
  normalized by the bare surface potential $V_0$ (see the discussion
  on the inset of Fig. \ref{fig:es}). The dashed line represents
  $\zeta = \zeta_\text{th}$ as given By Eq. (\ref{eqn:ampliNL}).\label{fig:zeta_robustesse}}
\end{figure}

We finally investigate the influence of electric parameters on $\zeta$
potential. Fig. \ref{fig:zeta_robustesse} presents a comparison
between the measured $\zeta$ potential and the slip prediction
\eqref{eqn:ampliNL}, for various electric parameters (with added
salt).
The excellent correspondence between the two quantities confirm the
robustness of the suggested picture, namely the $\zeta$ potential
originates in the coupling between electrostatic properties of the ion
cloud and the hydrodynamic behavior of the solvent, in the vicinity of
charged surfaces.

In the no-salt case, it is not possible to relate the streaming
current to the zeta potential only because the electric potential is not fully
screened over the fluid slab. However the simulations results are very well described
by a simple model which takes into account the electric properties at a PB
level and the hydrodynamic ones at a continuum level in the presence
of slip (results not shown).

\begin{figure}
\includegraphics[width=7cm]{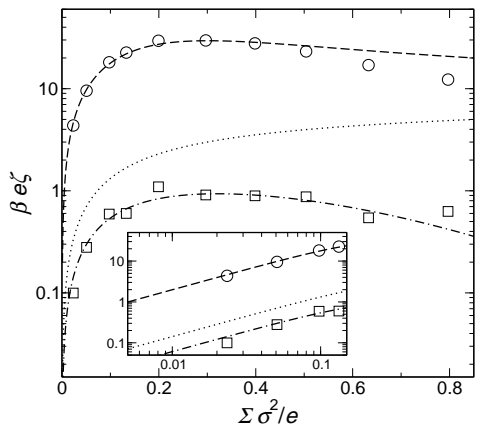}
\caption{Zeta potential as a function of surface charge
  $\varSigma$ ($\lb=\sigma$, $\rhos \sigma^3 = 0.06$). Symbols: MD
  results for the wetting 
  ($\scriptscriptstyle\square$) and the non-wetting ($\circ$)
  case. Dotted line: bare surface potential $V_0$, obtained from the
  PB prediction at a given $\kappa=1.3\sigma$ (see the discussion in
  the inset of Fig. \ref{fig:es}). Dashed line: slip prediction
  using the analytical fit of $b$ variation with $\varSigma$ in Eq. (\ref{eqn:b_full}) (see
  Fig. \ref{fig:b_charge} and associated text). Dashed-dotted line:
  no-slip prediction using the analytical fit of $z_\text{s}$
  variation with $\varSigma$ (see Fig. \ref{fig:zs_charge} and
  associated text). The inset is a zoom in the small $\zeta$ small $\varSigma$
  region.
  \label{fig:zeta_abs_charge}}
\end{figure}

We now focus on the role of surface charge $\varSigma$. Figure
\ref{fig:zeta_abs_charge} shows the evolution of the absolute
$\zeta$ potential with surface charge.
An important point is that it is possible to fully account for the
simulation results with analytical expressions: in the wetting case,
we simply used the value of the electric potential at the plane of
shear $V(z_\text{s})$, together with the analytical fit for
$z_\text{s}(\varSigma)$, derived in section \ref{sec:dynamique}; in
the non-wetting case, we combined the formula \eqref{eqn:ampliNL},
relating $\zeta$ to $b$ and electric parameters, with the analytical
expression for $b(\varSigma)$, \eqref{eqn:b_full} derived in section
\ref{sec:dynamique}. Moreover, we used the PB expression for the
evolution of electric parameters involved in $\zeta$ expression
($V(z)$, $V_0$, $\kappa_\text{eff}$) as a function of surface charge.

The non monotonous behaviour of $\zeta$ with $\varSigma$ is an interesting feature:
while the bare surface potential $V_0$ increases continuously with
$\varSigma$, the amplification factor -- related to the HBC --
decreases, following an increase of electric friction ($1/b'$). Those two
conflicting effects result in a non-monotonous evolution, and the
apparition of a maximum for $\zeta(\varSigma)$. This represents a
particularly important signature of the intimate coupling between
hydrodynamics and electrostatics in the definition of the $\zeta$
potential. We now discuss 
some consequences of
these results in the following section.

\section{Discussion} 
\label{sec:exp}

\subsection{$\zeta$ potential on hydrophobic surfaces : a slippage effect ?}

As we have shown above, the interplay between surface properties and electrokinetic phenomena
may lead to $\zeta$ potential much higher than the 
real
surface potential on hydrophobic surfaces, due in particular to the {modification of surface hydrodrodynamic properties and the onset of
liquid/solid slippage}.
In realistic experimental situations, the dynamical amplification
ratio $1+\kappa b$ can take important values, even in the case of a moderate 
slip
length : indeed, slip lengths are expected to range in the tens of nanometers
range, as predicted
theoretically \cite{barrat99b} and evidenced experimentally 
\cite{cecile05,lettre_exp}, while the Debye length $\ld$ is typically
of a few nanometers in quite standard aqueous electrolyte solutions 
(from $3$ \AA\ with a  1M salt, to $30$ nm at $10^{-4}$ M). A factor of $10$ 
 between $\zeta$ and $V_0$ is therefore quite conceivable.

This analysis therefore provides an interesting scenario to interpret the observation of
important $\zeta$ potentials on different hydrophobic
surfaces\,\cite{chibowski78,laskowski69,parreira61,ozon02,schweiss01}.
Despite the expected weaker surface charge, the measured $\zeta$
potentials are typically of the same order of magnitude as those
observed on hydrophilic surfaces\,\cite{scales92}. This could in fact
arise from small bare surface potentials and slip-induced
amplification.

We note however that another interpretation is usually suggested:
surface charge could be increased by preferential adsorption of anions
(OH$^-$ or Cl$^-$) at the wall. Recent AFM\,\cite{hu97} and
numerical\,\cite{kreuzer03} studies show that this phenomenon indeed
contributes to a certain extent to increase surface charge.

Finally we mention a recent experimental work by Churaev \etal
\cite{churaev02}, who performed $\zeta$ potential measurements for
aqueous KCl solutions in silanized quartz capillaries.
When a non-ionic surfactant is added to the solution, the measured
$\zeta$ potential decreases; this effect is attributed to
the disappearance of hydrodynamic slippage at the walls, resulting from
the surfactant adsorption. Using formula \eqref{eqn:ampli}, the
authors estimate the \emph{effective} slip length 
(averaged over the surface of the channel) to lie between
$\units{5}{nm}$ and $\units{8}{nm}$.
This first experiment using electrokinetics to characterize the dynamical
properties of the liquid at the interfaces is encouraging, 
but it seems that the results should be interpreted with care.
Indeed, to apply Eq. \eqref{eqn:ampli}, the authors assume that the
adsorption of the surfactant only affects wetting properties of the
surface (and therefore slip properties), electric properties
being unaffected, which is difficult to assess.


To overcome those obstacles, comparative static (using SFA or AFM)
and dynamic (with electrokinetics) measurements \emph{on the same
surfaces} would enable to test critically the relevance of
 the mechanisms suggested by this
work. To the best of our knowledge, no such experiment has already been
carried on.

\subsection{Zeta potential versus surface charge}

In light of the previous results,  it may be possible to
generate very large electrokinetic effects by using polarized
hydrophobic surfaces\,\cite{schasfoort99}. However, we saw in sections
\ref{sec:dynamique} and \ref{sec:ek} 
that eventually, too high surface charges will enhance friction and
therefore reduce the efficiency of hydrodynamic
slippage. Consequently, there should be an optimal choice for the
imposed surface charge, where the bare surface potential {reaches its maximum value compatible with a small contribution of electric terms to the friction, therefore leaving the slip length almost unaffected} (see Fig. \ref{fig:zeta_abs_charge}).

To determine the largest achievable zeta potential according to our
model, we now estimate the cross-over charge at which
the electric friction equals the LJ friction, which gives
$\varSigma_\text{c} \sim 0.3 e/\sigma^2$. For $\sigma =
\units{0.5}{nm}$, this value corresponds roughly to
$\units{0.2}{C/m^2}$. Furthermore, using $\lb = \kappa = \sigma$, we
find for the corresponding \emph{bare} surface potential $V_0 \sim
\units{80}{mV}$ (and therefore $\zeta \sim 10 V_0 \sim
\units{800}{mV}$).

This estimation for the optimal $\zeta$ potential is ten times larger
than typical values which can be obtained on hydrophilic surfaces
\cite{scales92}. This should motivate experiments to test this prediction.
We believe that the phenomena
discussed provide interesting hints concerning
 the possibility to generate
efficient electro-osmotic flows, with direct applications in
microfluidics.

\subsection{Curvature effects}

Our prediction for the dynamical amplification ratio, $1+\kappa b$ can take 
arbitrarily large values as the slip length $b$ goes to infinity (corresponding to 
a vanishing shear boundary condition at the interface). This prediction corresponds
however to a planar surface and we show in 
this paragraph that curvature effects actually lead to a saturation of the amplification ratio
for the zeta potential.

To demonstrate this effect, we shall generalize the calculation of electrophoretic
mobility of a sphere initiated by Smoluchowski\,\cite{smoluchowski} in 1921,
to take into account the partial slip BC at the sphere surface.

Let us consider a dielectric sphere, with radius $a$, immersed in an
electrolyte solution and submitted to an external electric field
$\bm{E}_\infty$.
We assume that the width of the EDL is small compared to the size of
the sphere ($\kappa a \gg 1$), and that the sphere permittivity is
small compared to the one of the liquid\,\footnote{This hypothesis
is reasonable for most of the particles in water, but  is  not
necessary: O'Brien and White have shown that the mobility is
independent of the
particle permittivity\,\cite{obrien78} (using a no-slip boundary condition).}.
Finally we consider that the external electric field does not modify
the ions distribution inside the
EDL\,\footnote{This hypothesis can be relaxed: O'Brien\,\cite{obrien83}
gave the complete solution (with a no-slip boundary condition).}.
We then solve the equations of the problem separately inside and
outside the EDL. We note $\mathcal{S}^+$ the sphere separating the EDL
and the exterior.

Since the sphere permittivity is very low, the normal component of the
electric field vanishes just outside the EDL. Inside the EDL, the
problem is reduced to an electro-osmosis one: neglecting the $1/r^2$ terms
in the Laplacian of the velocity field, we recover literally the
equations of the \emph{plane} case, simply replacing $v_t$ with
$v_\theta$ and $n$ with $r$\,\footnote{This approximation corresponds
to neglecting terms of order $\mathcal{O}(1/(\kappa a)^2$.}.
Thus we can immediately write:
\begin{equation}
\bm{v}_{\|} \Bigr|_{\mathcal{S}^+} =
- \frac{\ed\zeta}{\eta} \bm{E}_{\|} \Bigr|_{\mathcal{S}^+} .
\end{equation}
Finally, for an irrotational flow, the potential of velocities obeys
Laplace equation in the liquid outside the EDL, just as the
electric potential.  As the BCs at infinity and on $S^+$ are
identical, we can identify both fields at the constant
$\ed\zeta/\eta$. We directly deduce that $\bm{v}_\text{e} =
(\ed\zeta/\eta) \bm{E_\infty}$.

Nevertheless, the partial slip BC has to be modified to take into
account the sphere curvature\,\cite{einzel90}: indeed, this condition
accounts for the equality of friction $\lambda v_\theta$ and viscous
$\sigma_{r\theta}$ tangential stresses, but in spherical coordinates
the expression of the stress tensor at a radius $a$ includes a
curvature term: $\sigma_{r\theta} = \eta(\partial_r v_\theta -
v_\theta/a)$. We obtain a generalized slip condition at a surface with
curvature $a$:
\begin{equation}
\dep{v_\theta}{r} = v_\theta \left( \frac{1}{b_0} + \frac{1}{a}
\right) ,
\label{cl:gp_courbe}
\end{equation}
where $b_0=\eta/\lambda$ is the intrinsic slip length on a plane
surface; we define the slip length $b_\text{c}$ on a curved surface by:
\begin{equation}
\frac{1}{b_\text{c}} = \frac{1}{b_0} + \frac{1}{a} .
\label{eqn:b_courbe}
\end{equation}
This equation shows that
slippage is limited by the smallest of $b_0$ and $a$ and
curvature leads to a saturation of the effective 
slip length as $b_0 \rightarrow \infty$. 
Coming back to the calculation of the zeta potential,
the problem  is equivalent to that of  
a planar surface with slip length $b_\text{c}$. We therefore simply
deduce from Eq. \eqref{eqn:ampli} the effective $\zeta$ potential:
\begin{equation}
\zeta = V_0 (1+ \kappa b_\text{c}) .
\label{eqn:zeta_courbe}
\end{equation}

The zeta potential therefore depends on the ratio $a/b_0$, with the
limiting values : $\zeta_0=V_0 (1+ \kappa b_0)$ for $b_0 \ll a$,
independent of the radius of the particle; and $\zeta_c=V_0 (1+ \kappa a)$
in the opposite limit $b_0\gg a$. 

\section{Conclusion}

{The liquid properties close to an interface, embodied in the so-called Hydrodynamic Boundary Condition} is intimately related to the friction properties of
the liquid at the solid surface. In this paper we investigated
extensively the electric contribution to liquid/solid friction, and
how it affects the slip properties of liquids in the presence of
charged walls.
Various electric parameters were considered, and we provided a simple
but efficient model of electric friction, validated numerically thanks
to Molecular Dynamics simulations.
This model describes the slip length $b$ dependence with the surface
charge $\varSigma$ and the Bjerrum length $\lb$ of the
solvent. Moreover, it explains the weak dependence of $b$ with the
salt concentration measured in the simulations.

We then addressed the consequences of such a hydrodynamic slippage on
electrokinetic phenomena, through the coupling between
hydrodynamics and electric charge within the Electric Double Layer.
In this work, we extended a previous study \cite{lettre_ek}, by 
considering various electric parameters and electrokinetic
configurations. We confirmed the robustness of the suggested picture,
namely the widely used ``zeta potential'' -- characterizing the
amplitude of electrokinetic effects -- is not only a signature of
electrostatic interfacial features, but is also intrinsically
related to the dynamics of the solvent at the solid surface, providing
new perspectives to control this quantity. A similar conclusion was
reached in recent work \cite{dufreche05}.
In particular, we showed the existence of strongly amplified
electro-osmotic effects on hydrophobic surfaces through the induced
slippage, in quantitative agreement with previous streaming current
simulations.
In the slipping case, we discussed the non-trivial role of the surface
charge $\varSigma$ on the $\zeta$ potential: in addition to its direct
influence on the surface potential $V_0$, it enhances friction
and therefore reduces the efficiency of hydrodynamic slippage.
These conflicting effects lead to a non-monotonous
variation of $\zeta$ with $\varSigma$, which represents an important
signature of the coupling mechanisms suggested here.
Besides, the simulation results were shown to be in excellent agreement
with predictions taking into account the slippage of the fluid at the
solid surface. The amplification effect is accordingly controlled by
the ratio between the slip length (of the fluid at the solid surface),
and the Debye length.

Furthermore, practical consequences of our work have been discussed,
reconsidering existing experimental data of $\zeta$ potentials on hydrophobic surfaces,
and 
suggesting the possibility to generate strongly enhanced
electro-osmotic and electrophoretic effects in microchannels, using in
particular polarized hydrophobic surfaces. If confirmed, this feature would provide
various interesting applications for microfluidic devices.

Finally, this work suggests the possibility to use the zeta
potential as a new observable for the characterization of interfacial
hydrodynamics, through the coupling at small scales of electrostatics
and hydrodynamics.
In a following step, we plan to refine the model, with a more
realistic description of solvent
and ions  (SPC/E model of water, ions
with different sizes and wetting properties, etc.), in order to relate
in a more quantitative way electrokinetics measurements and slippage
properties.

\begin{acknowledgments}
The authors wish to thank Armand Ajdari for fruitful discussions.
\end{acknowledgments}

\appendix

\section{Zeta potential and slip}
\label{sec:zeta_glissant}

We detail here how the standard calculation of $\zeta$
potential \cite{hunter_zeta} is modified when a partial slip BC applies
at the walls, in both streaming current and electro-osmosis situations.

\subsection{Streaming current}

\begin{figure}
\includegraphics[width=5cm]{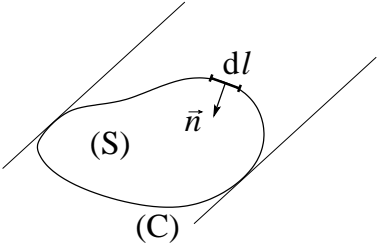}
\caption{Illustration of the channel used for $\zeta$
  calculation.\label{fig:canal_zeta}}
\end{figure}

To calculate the streaming current expression, we consider a channel
of arbitrary section (Fig. \ref{fig:canal_zeta}).
We assume that the Debye length $\ld$ is small compared to the typical
size of the channel $R$ and to the local inverse curvature of the channel $r$
everywhere on the contour $\mathcal{C}$. We can therefore consider
that the channel surface is flat at the EDL scale.

The elementary contribution to streaming current due to the
displacement of the EDL past a channel slice of width $\mathrm{d}l$
is:
\begin{equation}
\mathrm{d}I_\mathrm{e} = \mathrm{d}l \int_0^{\infty} \rhoe(n) v(n)
\mathrm{d}n ,
\label{eqa:di1}
\end{equation}
where $\infty \gg  \ld$ and $n$ is the coordinate normal to the
surface. We then substitute $\rhoe$ using Poisson equation:
\begin{equation}
\mathrm{d}I_\mathrm{e} = - \ed \mathrm{d}l \int_0^\infty \ddep{V}{n}
v(n) \mathrm{d}n .
\label{eqa:di2}
\end{equation}
To evaluate this integral, we must introduce the boundary conditions for
the velocity $v$ and the potential $V$ at the wall ($z=0$) and far
from the wall ($z=\infty$).
The velocity follows a partial slip BC: $v(0) = b\,\partial_n v$ (we
assume that $b \ll r$, so that curvature effects can be neglected
\cite{einzel90}); we note $V_0 = V(0)$ the electric potential at the
wall. Far from the wall, the potential is constant ($\partial_n V =
0$), conveniently fixed at $V(\infty)=0$.
We can now integrate by parts the expression in \eqref{eqa:di2}:
\begin{subequations}\label{eqa:i1}
\begin{align}
& \mathrm{d}I_\mathrm{e} = - \ed \mathrm{d}l \biggl\{ \left[\dep{V}{n}
v\right]_0^\infty - \int_0^\infty \dep{V}{n} \dep{v}{n} \mathrm{d}n
\biggl\}\\
& = - \ed \mathrm{d}l \biggl\{
- \underbrace{\left.\dep{V}{n}\right|_{0} v(0)}_{\sim \kappa V_0
  \bar{v}}
- \underbrace{\left[V \dep{v}{n}\right]_0^\infty}_{\sim \kappa V_0
  \bar{v}}
+ \underbrace{\cancel{\int_0^\infty V \ddep{v}{n} \mathrm{d}n}}_{\sim \kappa 
V_0
  \bar{v} \mathcal{O}(\ld/R)} \biggl\}\\
& = - \ed \mathrm{d}l \biggl\{ -\left.\dep{V}{n}\right|_{0} \times
b \left.\dep{v}{n}\right|_{0} + V_0 \left.\dep{v}{n}\right|_{0}
\biggl\}\\
& = - \ed \mathrm{d}l \times V_0 \left.\dep{v}{n}\right|_{0}
\biggl\{ 1 + b \frac{-\partial_n V (0)}{V_0} \biggl\} .
\end{align}
\end{subequations}
We then integrate this expression on the contour of the channel:
\begin{equation}
I_\mathrm{e} = - \ed V_0 \left( 1 + b \frac{-\partial_n V (0)}{V_0}
\right) \times \oint_\mathcal{C} \dep{v}{n} \mathrm{d}l .
\label{eqa:i2}
\end{equation}
The second Green theorem\,\cite{gradshteyn} enables us to transform
this expression:
\begin{equation}
I_\mathrm{e} = - \ed V_0 \left( 1 + b \frac{-\partial_n V (0)}{V_0}
\right) \times \iint_\mathcal{S} \Delta v\,\mathrm{d}S .
\label{eqa:i3}
\end{equation}
The Laplacian of velocity is directly related to forcing through
Stokes equation: $\Delta v = (1/\eta)(-\nabla p) = \text{Cst}$. We
finally obtain:
\begin{equation}
I_\mathrm{e} = - \frac{\ed V_0 ( 1 + \kappa_\text{eff} b)}{\eta}
\times \mathcal{A} \times (-\nabla p) ,
\label{eqa:i4}
\end{equation}
where the \emph{effective Debye length} $\kappa^{-1}_\text{eff}$ is
defined by $\kappa_\text{eff} = -\partial_n V (0)/V_0$, and goes to
$\ld$ in the linear limit $e V_0 \ll \kt$ ($\mathcal{A}$ is the area
of the channel section $\mathcal{S}$).
The $\zeta$ potential, derived from streaming current measurements, is
therefore related to the bare surface potential by a dynamical
amplification factor arising from slip:
\begin{equation}
\zeta =  V_0 ( 1 + \kappa_\text{eff} b) .
\label{eqa:i5}
\end{equation}
The effective length $ 1/\kappa_\text{eff} $ is discussed in section 
\ref{app:kappaeff}

\subsection{Electro-osmosis}

To calculate the electro-osmotic velocity on the same system, we
solve Stokes equation thanks to the BCs introduced in the preceding
sub-section:
\begin{equation}
- \eta \ddep{v}{n} = \rhoe E_\text{t} .
\label{eqn:stokes}
\end{equation}
We replace $\rho_e$ by its expression derived from Poisson equation:
\begin{equation}
\eta \ddep{v}{n} = \ed E_\text{t} \ddep{V}{n} .
\label{eqa:der2}
\end{equation}
This equation is integrated, assuming $\partial_n V$ and
$\partial_n v$ both vanish far from the wall:
\begin{equation}
\eta \dep{v}{n} = \ed E_\text{t} \dep{V}{n} .
\label{eqa:der1}
\end{equation}
From this equation, we deduce that $v(0) = b\,\partial_n v (0) = b
(\ed E_\text{t} / \eta) \partial_n V (0)$.
A second integration between the wall -- where $V=V_0$ and $v(0) =
b\,\partial_n v (0)$ -- and a plane located outside the EDL -- where
$V=0$ and $v = \text{Cst} = v_\text{eo}$ -- leads to:
\begin{subequations}\label{eqa:der0}
\begin{align}
v_\mathrm{eo} - v(0)& = -\frac{\ed E_\text{t}}{\eta} \Bigl( V_0 - 0
\Bigr)\\
v_\mathrm{eo}& = -\frac{\ed E_\text{t}}{\eta} \left( V_0 - b
\left.\dep{V}{n}\right|_{0} \right) ,
\end{align}
\end{subequations}
from which we extract the electro-osmotic velocity:
\begin{equation}
v_\mathrm{eo} = - \frac{\ed \zeta}{\eta} \times E_\text{t} ,
\label{zeta:EO_veo}
\end{equation}
with the $\zeta$ potential:
\begin{equation}
\zeta =  V_0 ( 1 + \kappa_\text{eff} b) .
\label{eqa:zeta}
\end{equation}

The fact that we get the same $\zeta$ potential for both phenomena
complies with the Onsager reciprocity theorem.

\subsection{The effective Debye length}
\label{app:kappaeff}

In section \ref{ssec:stream}, we introduced the length scale
$1/\kappa_{\text{eff}} = - V_0/\partial_n V(0)$ that allows to relate the
surface potential $V_0$, the slip length $b$ and the effective
$\zeta$ potential [see e.g. Eq. (\ref{eqn:ampliNL})].
In the weak overlap regime where the double layer is not affected
by the opposite boundary, one may use the analytical solution of PB
theory in the planar geometry \cite{hunter} to obtain
\begin{equation}
\kappa/ \kappa_{\text{eff}} \, = \, \hbox{argth}(\gamma) \,\frac{1-\gamma^2}{\gamma}
\end{equation}
where 
$\gamma = -\kappa \lgc + \sqrt{(\kappa \lgc)^2+1} $
and $\lgc$ denotes the Gouy length $\lgc = (2\pi \ell_B \varSigma)^{-1}$.
In the weak coupling regime (low $\Sigma$ where $\gamma$ vanishes),
we find $\kappa_{\text{eff}} \simeq \kappa$ while conversely, for
high charges, $\kappa/\kappa_{\text{eff}} \sim -\kappa \lgc \log(\kappa \lgc)$
which decreases as $\log(\Sigma)/\Sigma$.

The effective length $1/\kappa_{\text{eff}}$ is reminiscent of a related 
quantity $l_{GW} = -\partial_n V(0)/\partial^2_n V(0) $
introduced some time ago as a scaling length for charged interfaces
\cite{gueron}. Upon increasing the surface charge, $l_{GW}$
crosses over from the Debye length at low $\varSigma$ (as 
$1/\kappa_{\text{eff}}$) to the Gouy length. It therefore decreases
as $1/\varSigma$ for large interfacial charges, whereas $1/\kappa_{\text{eff}}$
is slightly larger [$\propto \log(\varSigma)/\varSigma$].

Finally,
it appears that the increase of $\kappa_{\text{eff}}$ at large 
$\varSigma$ due to non linear
effects is overruled by the decrease of  the slipping length
($b \propto 1/\varSigma^2$, see section \ref{ssec:ratio}): increasing $\varSigma$ 
decreases the slip-driven amplification factor $1+\kappa_{\text{eff}}\,b$, see Fig 
\ref{fig:zeta_abs_charge}.

\section{Wetting properties of ions}
\label{sec:wetting}

To obtain the present results, we considered a very simple model
with identical wetting properties for all liquid atoms, including ions.
This specific choice of LJ parameters aims at separating
electric properties at the interface from the dynamical slipping
behavior of the liquid.
To separate these effects, ions were therefore supposed to interact
with the wall in the same way as the solvent, so as to keep the
interfacial friction unaffected.
Moreover, no effect of salt on slip length has been reported
experimentally in the literature, therefore suggesting that no
specific behavior of the ions as compared to the solvent occurs at the
solid surface. In other words, solvent-surface and ion-surface
interactions are expected to be rather similar, and not very
asymmetric.
However, so as to explore the influence of our specific choice of LJ
interactions on the solvent dynamics, we performed complementary
simulations, 
exploring various solvent-surface ($c_{FS}$) and ion-surface ($c_{IS}$)
interactions, in the parameter window $c_{FS},c_{IS} \in [0.5;1]$
(with different values for $c_{FS}$ and $c_{IS}$), for different salt
concentrations. We calculated both equilibrium and dynamical quantities
for these parameters.

\begin{figure}
\includegraphics[width=7cm]{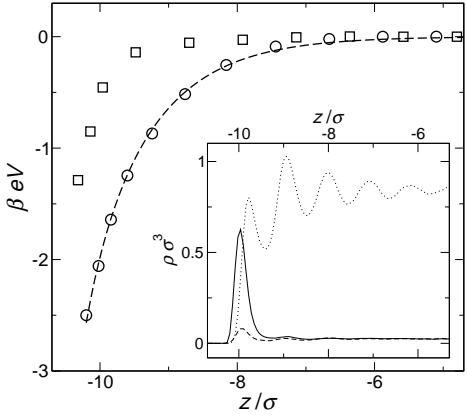}
\caption{Electrostatic potential for a non-wetting solvent
  ($c_{FS}=0.5$) and ions with different wetting properties
  ($\rhos\sigma^3=0.07$). Dashed line: bare PB prediction; Symbols: MD
  results for $c_{IS}=0.6$ ($\circ$) and $c_{IS}=1$
  ($\scriptscriptstyle\square$). Inset: density profiles in the
  strongly asymmetric case ($c_{FS}=0.5$ and $c_{IS}=1$): solvent
  ($\cdots$), counter-ions (---) and co-ions
  ($--$).\label{fig:es_cis}}
\end{figure}

Let us first focus on the equilibrium properties of the EDL.
For moderately
asymmetric situations ($\delta c_{IS} \lesssim 0.2$), the modified PB
approach (Sec. \ref{sec:es}) remains valid. Indeed, as can be seen on
Fig. \ref{fig:es_cis}, the standard PB
potential (without structuration) describes quite well the electric
potential, computed from the simulation by double-integration of 
the charge density  profile $\rhoe(z) = e(\rho_+(z) - \rho_-(z))$.
Nevertheless, for strongly asymmetric situations -- typically with a
``non-wetting'' solvent ($c_{FS}=0.5$) and ``wetting'' ions
($c_{IS}=1$) --, we observe specific ions-wall interactions, and the
modified PB model fails to describe the electric potential at the
interface (see Fig. \ref{fig:es_cis}).
To conclude on the static properties, the modified PB approach, even
if it fails to describe specific ions-solid interactions, provides an
accurate description for moderately asymmetric parameters, which we
think are representative of experimental situations.

As far as dynamical properties are concerned, all the situations considered confirm
the robustness of the picture put forward in the main text. On
Fig. \ref{fig:zeta_robustesse}, we compare
the $\zeta$ potentials measured from the simulations with the slip
prediction \eqref{eqn:ampliNL}.
For moderately asymmetric situations,
we used the PB potential as an input, whereas for strongly asymmetric
ones we used the ``exact'' surface potential, extracted from the
simulations. These quantities correspond to each other perfectly.
Thus these complementary results confirm that the message emerging
from the simpler case of equal LJ interactions is not modified when
exploring a larger parameter space: namely the zeta
potential is a signature of the coupling between the electrostatics and the
fluid dynamical properties at the surface.

\section{Density functional theory and the modified Poisson-Boltzmann
equilibrium}
\label{app:C}

For arbitrary solvent and microionic density profiles
[resp $\rho_f({\bm r})$ and $\rho_{\pm}(\br)$], the total free energy 
of our confined system may be written as the functional
\begin{eqnarray}
\beta {\cal F}\{\rho_f,\rho_\pm\} &=&
\sum_{i \in f,+,-} \int_{\mathcal{V}} \rho_i(\br) 
\left[ \log\left(\Lambda_i^3 \rho_i(\br) \right)
-1\right]  d\br
\nonumber\\
&+& \beta  {\cal F}^{\text{LJ}}_{\text{excess}}\{\rho_{\text{tot}}\}
+ \beta  {\cal F}^{\text{Coul}}\{\rho_+,\rho_-\},
\label{eq:fonc}
\end{eqnarray}
where $\mathcal{V}$ denotes the total volume and the $\Lambda_i$
are irrelevant length scales.
The first term on the rhs acounts for the ideal gas entropy while the
excess part is included in $ {\cal F}^{\text{LJ}}_{\text{excess}}$.
This {\it a priori} unknown functional depends on the total density
only 
\begin{equation}
\rho_{\text{tot}}=\rho_f+\rho_++\rho_-.
\label{eq:foot}
\end{equation}
This is a consequence of
treating solvent and microions on equal footings, not only in their
mutual interactions, but also in their interaction with wall atoms.
Finally, the last term on the rhs of Eq. (\ref{eq:fonc}) denotes the
Coulombic contribution to the total free energy. Unlike its Lennard-Jones
counterpart, treated exactly (but formally), 
$ {\cal F}^{\text{Coul}}\{\rho_+,\rho_-\}$ is approximated,
in a mean-field spirit (see e.g. \cite{pre2000}), by 
\begin{equation}
\beta {\cal F}^{\text{Coul}}\{\rho_+,\rho_-\} \,=\, 
 \frac{1}{2} \int_{\mathcal{V}\times\mathcal{V}} \rho_e (\br) G(\br,\br') \rho_e(\br') d\br d\br',
\end{equation}
where $G(\br,\br')$ denotes the Green's function inverting the Laplacian
in the geometry under study and $\rho_e(\br)=e[\rho_+(\br)-\rho_-(\br)]$
is the local charge density, outside the confining walls.

Minimizing the total free energy $\cal F$ with respect to solvent density
leads to 
\begin{equation}
\log[\Lambda_s^3 \rho_s(\br)] + \frac{\delta \beta {\cal F}^{\text{LJ}}_{\text{excess}}}{\delta \rho_{\text{tot}}(\br)} \,=\,\hbox{cst}
\label{eq:b}
\end{equation}
both in the canonical ensemble where the total number of solvent
molecules is fixed, or in a grand canonical description where the chemical
potential is given. Here, explicit use was made of Eq. (\ref{eq:foot}).
On the other hand, minimization with respect to microionic densities 
yields
\begin{equation}
\log[\Lambda_\pm^3 \rho_\pm(\br)] + \frac{\delta \beta {\cal F}^{\text{LJ}}_{\text{excess}}}{\delta \rho_{\text{tot}}(\br)}  
\pm \beta e \int_{\mathcal{V}} \rho_e(\br') G(\br,\br') d\br' 
\,=\,\hbox{cst}'.
\label{eq:a}
\end{equation}
Inserting Eq. (\ref{eq:b}) into (\ref{eq:a}) and realizing that the electric potential
reads
\begin{equation}
\beta e V(\br) \,=\,  \int_{\mathcal{V}} \rho_e(\br') G(\br,\br') d\br' ,
\end{equation}
we obtain
\begin{equation}
\rho_\pm(\br) \, \propto \, \rho_s(\br) e^{\mp \beta e V(\br) }.
\label{eq:mpb}
\end{equation}
This is precisely the form obtained in Eq. (\ref{rhopm}) on more
heuristic grounds. Upon neglecting the LJ excess free energy, 
one recovers the standard
Poisson-Boltzmann relation.

The fact that the Lennard-Jones free energy functional only depends
on the total density $\rho_{\text{tot}}$ given by (\ref{eq:foot}) plays 
a pivotal role in the derivation of (\ref{eq:mpb}) and results from the
symmetrical role played by ions and solvent molecules as far as 
non ionic interactions are concerned. Whenever the wetting 
properties  of the ions differ from those of the solvent, 
$ {\cal F}^{\text{LJ}}_{\text{excess}}$ no longer depends on 
$\rho_{\text{tot}}$ but in general separately on $\rho_s$, $\rho_+$ and
$\rho_-$. A similar remark applies if the non Coulombic 
solvent-solvent interaction differs from solvent-ion and ion-ion
interactions. In those situations, Eq. (\ref{eq:mpb}) does not hold,
see e.g. Appendix \ref{sec:wetting}. 
While analytical progress might be possible,
we did not attempt a density functional description in these cases. 

{\em A criterion for the validity of the modified Poisson-Boltzmann
description (\ref{eq:mpb})}.
Equation (\ref{eq:fonc}) assumes a mean field factorization 
of the Coulombic energy, which discards microionic correlations. 
The importance of these correlations is quantified by the plasma
parameter $\Gamma = z^2 \ell_B/d_{ii}$ where $d_{ii}$ denotes
the typical distance between counterions of valency $z$ 
in the electric double layer \cite{levin02}. This distance
is most conveniently estimated assuming that the counterions
form a Wigner crystal at the planar interface, which gives
$d_{ii}^2 \varSigma = z e$. We consequently obtain the following
criterion for the validity of the mean-field assumption 
underlying the modified Poisson-Boltzmann picture :
\begin{equation}
\Gamma = z^{3/2} (\varSigma \ell_B^2/e)^{1/2}<1.
\label{eq:crit}
\end{equation}
{\it A priori}, the threshold value of $\Gamma$ 
is expected to be of order 1, and a survey of the literature
indicates that it is very close to unity, hence Eq. (\ref{eq:crit}) 
(see \cite{levin02,levin}  and
references therein). At fixed $\ell_B=\sigma$, (\ref{eq:crit})
implies for monovalent microions $\varSigma \sigma^2/e<1$,
which is always fulfilled in the situations investigated in the
present study. We emphasize here that the short distance
repulsion of the Lennard-Jones potential used ensures that
$d_{ii}>\sigma$, so that one always has $\Gamma = \ell_B/d_{ii}<1$
for $\ell_B=\sigma$ (and more generally for $\ell_B<\sigma$). 
On the other hand, considering $\varSigma = 0.2e/\sigma^2$,
(\ref{eq:crit}) translates, again for $z=1$, into $\ell_B <2.2 \sigma$.
We thus expect here deviations from the modified PB equilibrium
for $\ell_B>2.2 \sigma$. This point is discussed at the end
of section \ref{sec:es}.

\end{document}